\documentclass[conference,10pt]{IEEEtran}
\IEEEoverridecommandlockouts
\input{config.sty}
\usepackage[show]{chato-notes}
\usepackage{booktabs}
\usepackage[cm]{fullpage}
\usepackage{amsmath}
\usepackage{longtable}
\usepackage{tabularray}
\usepackage{array}
\usepackage{caption}
\usepackage{subcaption}
\usepackage{url}

\usepackage{geometry}
\IEEEoverridecommandlockouts
\begin{document}
\title{On Dual Connectivity in 6G Leo Constellations}
\author{Achilles Machumilane,~\IEEEmembership{Student Member,~IEEE,}
~Alberto Gotta,~\IEEEmembership{Member,~IEEE}\\
Institute of Information Science and Technologies (ISTI), CNR, Pisa. \\
e-mail:\{name.surname\}@isti.cnr.it}

\maketitle

\begin{abstract}
Dual connectivity (DC) has garnered significant attention in 5G evolution, allowing for enhancing throughput and reliability by leveraging the channel conditions of two paths. However, when the paths exhibit different delays, such as in terrestrial and non-terrestrial integrated networks with multi-orbit topologies or in networks characterized by frequent topology changes, like Low Earth Orbit (LEO) satellite constellations with different elevation angles, traffic delivery may experience packet reordering or triggering congestion control mechanisms. Additionally, real-time traffic may experience packet drops if their arrival exceeds a play-out threshold.
Different techniques have been proposed to address these issues, such as packet duplication, packet switching, and network coding for traffic scheduling in DC. However, if not accurately designed, these techniques can lead to resource waste, encoding/decoding delays, and computational overhead, undermining DC's intended benefits.
This paper provides a mathematical framework for calculating the average end-to-end packet loss in case of a loss process modeled with a Discrete Markov Chain - typical of a wireless channel - when combining packet duplication and packet switching or when network coding is employed in DC. Such metrics help derive optimal policies with full knowledge of the underlying loss process to be compared to empirical models learned through Machine Learning algorithms.

\end{abstract}


%
\IEEEpeerreviewmaketitle


\section{Introduction}

\IEEEPARstart{N}{on}-Terrestrial Networks (NTNs) is a novel communication framework introduced in the Third Generation Partnership Project (3GPP) standard from release 16 \cite{3gpp16}to extend the coverage of the mobile network through the space segment, thanks to the remarkable progress made with the deployment of the massive Low Earth Orbit (LEO) constellation with global coverage and propagation delays comparable to terrestrial trunks. The actual market take-up of 3GPP-NTN will probably take place with the new releases and will coincide fairly closely with the sixth generation (6G). However, many topological issues in the space segment still need to be investigated, tested, and demonstrated, as its deployment is more challenging compared to terrestrial networks. One of the most obvious differences is the highly dynamic evolution of the LEO network topology, which, even when it is deterministic, is constantly changing according to the relative movement of the satellites, with very short satellite visibility periods on the intra-satellite links (ISLs) and from the user equipment (UE) on the ground.
The two effects, i.e. \textit{i}) ground accessibility period and \textit{ii}) ISL availability time leads to a fragmentation of time into very short slots, generally referred to as "snapshots", during which both the access and the network topology can be considered unchanged. In \cite{cassara2022orbital}, the mean duration of the snapshots is measured by emulating some constellations of different sizes, using Starlink as a reference. The average snapshot duration in the order of a few minutes introduces continuous handovers between the satellites, which never occur in terrestrial mobile networks where the RAN and the associated gNB are fixed and only the UEs move at a very low speed concerning RAN. In addition, topology updates lead to constant changes in the routing tables when multi-hops are considered within the LEO constellation. As a result, data plane traffic can face constant fluctuations in channel latency and out-of-sequence (OoS) flows \cite{carpa2017evaluating}, leading to reordering issues. Packet reordering has a severe impact on much of the traffic on the Internet, such as Transport Control Protocol (TCP), Quick User Datagram Internet Connection (QUIC), and Real Time Protocol (RTP), causing throughput degradation (with TCP timeouts), retransmissions (in QUIC) and frame glitches due to OoS packet loss (with RTP flows). In networks with a higher degree of packet reordering, QUIC traffic does not collapse the congestion window until persistent congestion is declared as given in [RFC 9002] \cite{rfc9002}, unlike TCP, which collapses the congestion window upon the expiration of a retransmission time out (RTO)\cite{celandroni2006long, celandroni2006networking}. Instead of collapsing the congestion window and declaring everything in-flight lost, QUIC allows probe packets to exceed the congestion window when the timer temporarily expires. 

Streaming applications and protocols also suffer from reordering problems such as playout buffering for non-real-time services, dropout of out-of-sequence packets with glitches, or loss of the entire sequences of blocks of frames.
In \cite{bacco2022air}, the impact of the de-jitter buffer in RTP-based live streaming with multi-connectivity on real 3G measurements is analyzed: The de-jitter buffer forwards the first received copy of each packet (using the serial number). However, different paths are likely to result in different latencies which, considering the maximum tolerated jitter threshold, lead to a non-null probability of exceeding such a value.

But even when reordering does not occur, jittering of the packet pacing in VoIP applications results in a significant downgrade of the quality of experience (QoE) perceived by the end user. This paper investigates the mechanisms provided by 3GPP-NTN to mitigate the packet losses linked to the reordering problem with DC in NTNs and the related packet switching, packet splitting (PS), packet duplication (PD), and network coding (NC) techniques. Specifically, we provide a mathematical analysis to show the efficient way of using NC and combining PD and PS to mitigate packet dropouts because of the delay jitter in the two DC connections.  

\section{Dual Connectivity} \label{sec:dual_connectivity}
Dual connectivity (DC) is a technology in which a UE connects simultaneously to two independent access nodes (eNB or gNB) connected via an X2 or Xn interface\cite{3gpp12}. One node acts as a Master Node (MN) and the other as a Slave Node (SN) with at least the MN connected to the core. 3GPP introduced DC in Release 12 \cite{3gpp12} as a general term for multi-connectivity in LTE. From Rel 16, 3GPP has defined several new DC solutions for the NR to support NTN \cite{3gpp16}. In these DC architectures, a UE reaches the 5GCN either via an NTN-based NG-RAN and a cellular NG-RAN or via two NTN-based NG-RANs simultaneously. NTN platforms such as satellites can be applied with transparent and regenerative payloads with gNB or gNB-DU (Distributed Unit) function on the satellite's board. With the transparent payload, a satellite acts like an RF relay; in contrast, a regenerative satellite can demodulate, decode, re-encode, re-modulate, and filter an uplink RF signal before transmitting it on the downlink. The following describes the different 3GPP DC architectures with NTN-based NG-RAN.

\subsection{DC with transparent NTN-based NG-RAN}
\begin{figure}
\centering
\includegraphics[width=1\columnwidth]{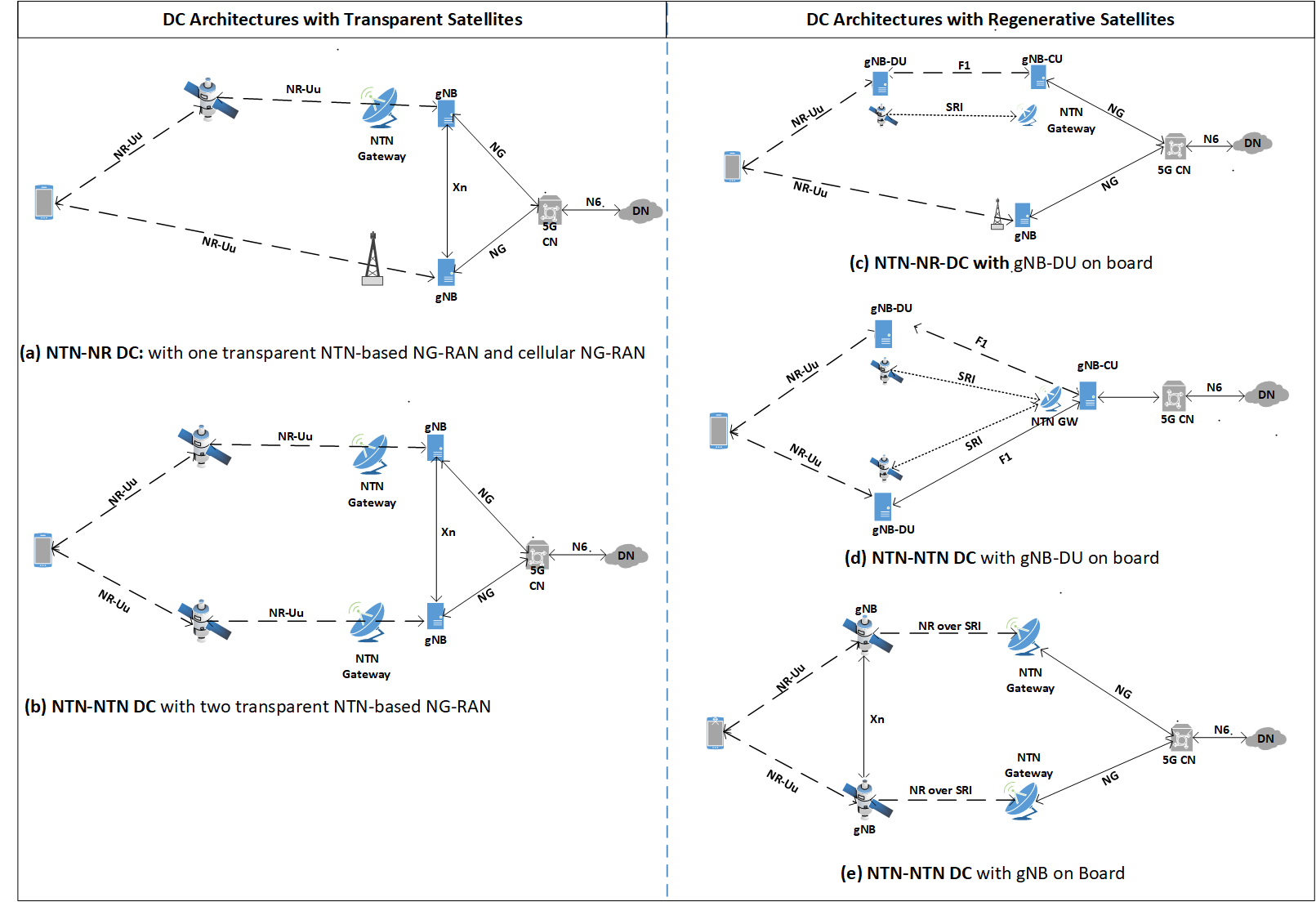}
\caption{NTN-Cellular based DC Architectures}\label{NTN-NR-DC-combined}
\end{figure}

This DC implementation involves one or two transparent satellites and two gNBs on the ground. 
Two configurations can be distinguished: \textit{(i)} NTN-NR DC architecture in which a UE is simultaneously connected to a 5GCN via a transparent NTN-based NG-RAN and a cellular NG-RAN, as shown in Figure \ref{NTN-NR-DC-combined}(a), with two gNBs located on the ground, one acting as an MN and the other as an SN, connected via an Xn interface. The PDCP, which is responsible for traffic distribution and aggregation, is deployed in the MN. \textit{(ii)} NTN-NTN DC architecture which involves two transparent NTN-based NG-RANs, as shown in Figure \ref{NTN-NR-DC-combined}(b). The two satellites can be GEO or LEO satellites or a combination of both, but the difference in RTT can cause the traffic carried over the two links to experience different delays.

\subsection{DC with regenerative NTN-based NG-RAN}
In DC with regenerative NTN-RAN, the satellites have gNB capabilities, either with the entire gNB or only the gNB-DU on board, while the central unit (CU) is deployed in the terrestrial segment. Three architectures are proposed: \textit{(i)}
    NTN-NR DC with gNB-DU onboard as shown in Figure \ref{NTN-NR-DC-combined} (c) consisting of a satellite with gNB-DU on board, which is connected to the gNB-CU via F1,  transported via SRI. The second connection is via the NR-gNB on the ground and the PDCP is implemented in the 5GCN; \textit{(ii)} NTN-NTN DC with gNB-DU onboard consists of two regenerative satellites with gNB-DU on board connected to the common gNB-CU on the ground via F1 interface transported over SRI, and \textit{(iii)} NTN-NTN DC with gNB onboard in which two satellites with gNB on board are connected via an Xn interface, one gNB with PDCP acting as a MN  and the other as a SN.

\subsection{Delay Impact}
The introduction of the NTN segment in the NG-RAN may cause the traffic traversing the two links to experience different delays depending on the architecture used, resulting in packet reordering and dropouts if the delay exceeds a threshold. The additional delay may be due to the difference between the RTT of the satellite and the NR link, as well as the additional interfaces involved in the RAN, such as Xn and F1. In cases where two satellites are involved, there may be a delay difference if the two satellites are different, e.g., GEO and LEO. Since the LEO link has a relatively low latency, it can be used for delay-sensitive traffic, and the GEO link for high throughput applications. If the two satellites are connected via an Xn interface, e.g., in an NTN-NTN DC with a gNB on board, the two satellites must be close to each other and have at least partially overlapping coverage areas. In such cases, combining GEO and LEO satellites is not feasible, as the Xn interface would cause long delays due to the different orbital planes and the distance between the two satellites. In general, the DC implementation requires adaptations to support the radio access technology with extended latency, variable latency within the backhaul network (e.g., an Xn interface), and the delay skew between the two involved radio access technologies.

To address DC-related challenges, 3GPP has proposed PS, PD, and NC as possible solutions \cite{3gpp3}. In the following, we introduce these techniques and investigate how efficiently they can be used to mitigate the delay problem in DC with heterogeneous connections characterized by different delays and resulting into packet losses.


\section{Packet Splitting, Packet duplication, and network coding}

\subsection{Packet Splitting}
Packet Splitting (PS) is the procedure that divides the traffic of a single data flow into several parts and distributes the parts among several access networks (paths). When PS is applied to a flow, part of the flow is transmitted over one access network and another part of the same flow is transmitted over another access network. PS is therefore suitable for load balancing across different access networks or paths and can increase data rates but does not provide any protection. The main limitation of PS is that it can introduce reordering due to delay differences between the paths and discarding of packets if the delay exceeds the threshold. Another design challenge is the split ratio which should take into account the path conditions in order to achieve the intended benefits.

\subsection{Packet Duplication}
Packet Duplication (PD) in DC allows delivering the same packet over two connections simultaneously to increase reliability and goodput. 3GPP has defined a PDCP Duplication Decision (PDD) function to handle PD in DC. The PDD can be deployed either in the CN or MN \cite{3gpp17}. PDD duplicates the traffic at the transmitter while at the receiver, it aggregates the traffic and discards the duplicates. PD can be used to support Ultra Reliable Low Latency Communications (uRLLC), which require high reliability. It can increase robustness to errors and fault tolerance as traffic can be delivered over one link if the other link fails. It can also solve the problem of packet buffering and reordering at the receiver and packet dropout due to delays exceeding the threshold. The most significant limitation of PD is that it reduces the information rate due to double transmissions. To address this problem, 3GPP has introduced a dynamic control option in Release 16 and Release 17, where PD can be adaptively enabled or disabled depending on the link conditions or the availability of network resources.  In \cite{machumilane2022actor}, authors propose a framework based on the Actor-Critic RL algorithm with dynamic PD based on the end-to-end packet loss rate threshold.

\subsection{Network Coding}
Network coding (NC) is a data coding technique that combines packets before transmission to increase robustness against network impairments \cite{li2003linear}. With NC, messages are not transmitted individually but are combined using algebraic algorithms and forwarded to the destination. The receiver adopts similar algorithms to decode the data. NC reduces the required number of transmissions but increases the computing load at the intermediate and destination nodes. It can also improve reliability, throughput, efficiency, and scalability and protect data from network attacks like eavesdropping. One of the NC techniques is Random Linear Network Coding (RLNC), in which random linear combinations of packets achieve coding by multiplying them by linear coefficients randomly selected from a finite field or Galois field (GF). Since RLNC uses random coefficients, it is suitable for networks with unknown or dynamic topology such as LEO satellite constellations. Compared to traditional routing, RLNC provides fully distributed computation and increased redundancy. It is easy to implement and reduces packet retransmissions in erasure networks. RLNC can help to reduce latency and increase efficiency and network capacity in wireless and satellite networks. 


\section{Problem Formulation}
\begin{figure}
\centering
\includegraphics[width=0.8\columnwidth]{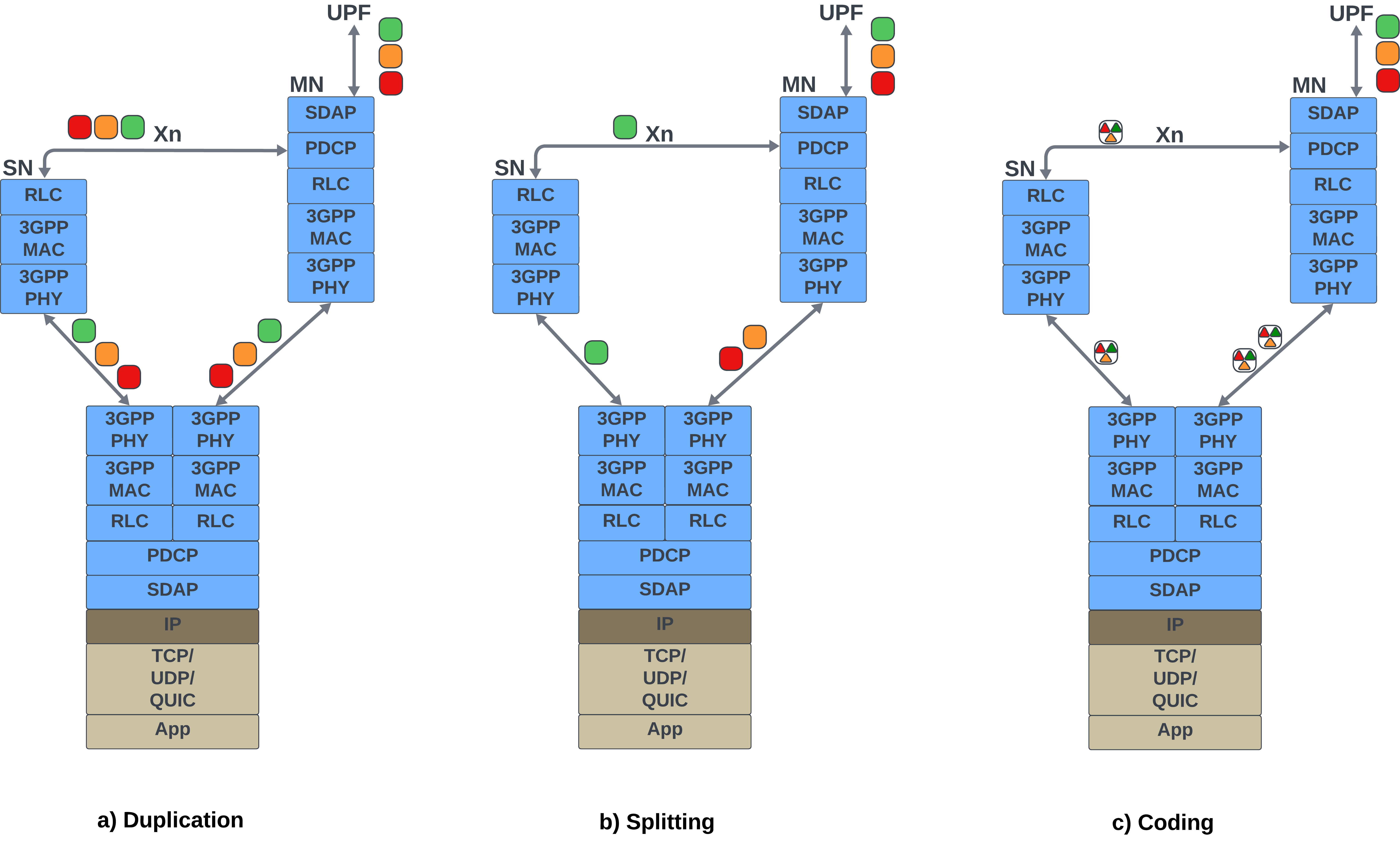}
\caption{Dual Connectivity Scenarios} \label{dc_scenarios}
\end{figure}

In this section, we formulate the problem to define the probability of losing packets on a flow of streamed packets using the DC architectures described in section \ref{sec:dual_connectivity}. In our investigated scenario, the layer devoted to managing the flow of packets is the PDCP. The PDCP receives service data units (SDUs) as IP packets and encapsulates them into a sequence of PDCP PDUs. Vice versa, the sequence of data packets from the UPF is transmitted to the UE by the MN according to the DC policy shared with the SN. The PDCP at the UE reconstructs the sequence of the PDCP PDU before sending it to the SDAP. Figure  \ref{dc_scenarios} shows how PDCP PDUs are delivered using different paths applying PD, splitting, and coding. 

The paths adopted in the analysis of this work are based on wireless channels whose behavior can be approximated through a two-state discrete-time Markov Chain (DTMC) model \cite{lutz1991land}, with a line-of-sight (LOS) state as a good state (G) characterized by a Ricean fading with high received signal power, and a blocking state as a bad state (B) characterized by Rayleigh fading, which represents areas with No-LOS, which causes erasures. Channel loss evolution with an underlying DTMC model is fully characterized by its transition matrix as follows:
\begin{equation}
\nonumber
T=\left(\begin{array}{cc}
     g & 1-g  \\
     1-b & b 
\end{array}\right)
\end{equation}
whose stationary probabilities $\pi_i$ are respectively:
\begin{gather}
\nonumber
\pi_G=\frac{g}{g+b},\:\:
\nonumber
\pi_B=\frac{b}{g+b}.
\end{gather}
Hence, the packet loss for the channel $i$ is given by $p_{loss_i}=\pi_{B,i}$.
Exploiting Eq. (3) $P_n(l|\cdot)$ in \cite{celandroni2011performance}, the probability that $l$ packets out of $N$ are lost on  channel $i$ is given by:
\begin{equation}
    P_{loss_i}(l|N)=P_n(l| N, b_i, g_i, \pi_{G,i}, \pi_{B,i})
    \label{eq:loss_pck_n_N}
\end{equation}

\subsection{Packet Duplication}

The first case analyzed is the packet duplication (DP) shown in Figure \ref{dc_scenarios} (a). In this case, a packet is lost iff both paths experience a loss, and the joint packet loss probability is given by:
\begin{equation}
   P^{(PD)}_{loss}=p_{loss_1}\cdot p_{loss_2}
   \label{eq:2-mp_Pbad}
\end{equation}
\subsection{Packet Splitting}

The second case analyzed is packet splitting (PS) shown in figure \ref{dc_scenarios} (b). In this case, a portion of packets $w_1 \cdot N=N_1$ is sent on path $1$ and another portion $w_2 \cdot N= N_2 $ on path $2$, with $w_1+w_2=1$, in a round of $N$ transmission. We assume that the two paths lose $l$ packets out of $N$. This means that $l_1$ packets are lost on path 1 out of $N_1$ and $l_2$ on path 2 out of $N_2=N-N_1$. In our case, we are interested in calculating 
\begin{equation}
\label{eq:sp_1}
P^{(PS)}_{loss}(l|N)=P_{loss}(l_1+l_2=l|N_1+N_2=N)
\end{equation}

which means to calculate  
\begin{equation}
\label{eq:sp_2}
P_{loss}(l_1,l_2|N_1,N_2)=\frac{P_{loss}(l_1,l_2,N_1,N_2)}{P(N_1,N_2)}
\end{equation}

Assuming that $p_{loss_1}$ and  $p_{loss_2}$ are independently distributed, the joint probability can be calculated as follows:
\begin{multline}
\label{eq:sp_3}
P^{(PS)}_{loss}(l|N)=\\
=\frac{P_{loss_1}(l_1,N_1,N_2)\cdot P_{loss_2}(l_2,N_1,N_2)}{P(N_1,N_2)}=\\
=P_{loss_1}(l_1|N_1,N_2)\cdot P_{loss_2}(l_2|N_1,N_2)\cdot P(N_1,N_2),
\end{multline}

by exploiting Bayes' theorem.
Given that the amount of packets $N_1$ and $N_2$ are deterministically assigned over the two paths \eqref{eq:sp_3} can be written as:
\begin{multline}
\label{eq:sp_6}
P^{(PS)}_{loss}(l|N)=P_{loss_1}(l_1|N_1)\cdot P_{loss_2}(l_2|N_2)=\\
=P_{loss_1}(l_1|N_1)\cdot P_{loss_2}(l-l_1|N-N_1).
\end{multline}

\subsection{Network Coding}
The last case that we analyze is the packet splitting with NC as shown in figure \ref{dc_scenarios} (c). This case is almost the same as the splitting, apart from the fact that the transmission reliability of a generation is increased by encoding them into $N$ packets.   
In this case, the PDCP protocol receives service data units (SDUs) as IP packets and organizes them into a source block of length $K$, the so-called generation. Such generation is processed by the RLNC function to produce $N$ coded packets encapsulated into a sequence of coded PDCP PDUs. The coded PDCP PDUs are enumerated using the same PDCP sequence number and then transmitted using different paths. The receiver collects the coded content from all incoming PDCP PDUs tagged with the same sequence number coming from the different paths until the content of the source block is reconstructed. 

As proven by equations (\ref{eq:sp_1}-\ref{eq:sp_6}) the loss probability of $l$ packets given that $N$ packets are transmitted, assuming that the packets are split into two sub-flows $N_1$ and $N_2$ each one experiencing a packet loss of $l_1$ and $l_2$, respectively, is still given by \eqref{eq:sp_6}, i.e., $P^{(NC)}_{loss}(l|N)=P^{(PS)}_{loss}(l|N)$.

However, for the RLNC, we recover the information iff at least $K$ packets are correctly received, and the decoding procedure doesn't fail, that is:
\begin{multline}
\label{eq:loss}
P^{(NC)}_{rec}(N,K,q)=\\
=\displaystyle ( 1-\sum_{l=N-K}^N P^{(NC)}_{loss}(l|N))\cdot P^{(NC)}_{dec}(N,K,q),
\end{multline}

where $P^{(NC)}_{dec}(N,K,q)$ is the decoding probability exactly calculated in \cite{6188495} with Eq. (5) for a generation of $K$ packets out of $N$ over a Galois Field of size $q$ that is:
\begin{multline}
\label{eq:loss}
P^{(NC)}_{dec}(N,K,q)=\\
=\displaystyle \frac{\sum _{j=0}^{N-K} (-1)^j \binom{N}{j} \prod _{i=0}^{K-1} \left(q^N-q^i\right)}{\left(q^K-1\right)^N}
\end{multline}
Finally, the generation of $K$ information packets is lost with probability:
\begin{equation}
\label{eq:loss}
P^{(NC)}_{loss}(N,K,q)=\frac{K}{N}(1-P^{(NC)}_{rec}(N,K,q))
\end{equation}


\section{Performance Evaluation}

We evaluated our model considering a UL transmission in a  DC mode, where a mobile UE connects to two LEO satellites simultaneously. With the channel model described above and the channel parameters provided by ITU for the earth-space link at 2.2 GHz in dense urban areas \cite{recommendation2017propagation}, we calculated the stationary channel end-to-end packet loss rate (E2E-PLR) according to \cite{machumilane2023towards} and used the equations (\ref{eq:sp_6}-\ref{eq:loss}) to investigate how our model can overcome E2E-PLR with PS combined with PD, and NC. The evaluation results are shown in Tables I, II, and III. We considered three scenarios with a pair of satellites, $S1$ and $S2$, at different elevation angles, $70^{\circ}$-$60^{\circ}$ , $60^{\circ}$-$45^{\circ}$  and $70^{\circ}$-$45^{\circ}$. We used different parameters in the performance evaluation to investigate the trade-offs that can be considered when combining the three methods. The first row in all tables shows the load balance $LB$ ratio applied between $S1$ and $S2$. The second row shows the balance $DT$ between PD and PS when PD and PS are combined. The third row shows the average redundancy factor $RF$ or redundancy rate and the fourth row shows the E2E-PLR calculated by our model with different parameters. We compare the E2E-PLR achieved with PD and PS with that achieved with NC, varying parameters such as $K$, $N$, $LB$, and $DT$. 

The results show that NC can achieve a lower E2E-PLR compared to PD and PS. However, this comes at the price of higher encoding and decoding complexity, which reduces the information rate and increases delay. For example, looking at Table I when the links are at $70^{\circ}$-$60^{\circ}$, if an E2E-PLR of 0.005 is targeted, with PD and PS we must either transmit with a single transmission on satellite A, which is at a higher elevation angle and has a lower loss rate, or use both links with a higher redundancy of about 1.8, which reduces the information rate. In the $60^{\circ}$-$45^{\circ}$ scenario, the 0.005 could not be achieved by PS and PD even with $RF=1.8$, because the two links have higher loss probabilities. NC, on the other hand, could achieve the target with $RF=2.5$ when $N=10$ and $RF=2$ when $N=20$. With $N=10$, the NC can achieve an E2E-PLR of up to 0.001\% with $RF=5$. When $N$ is fixed, decreasing $K$ increases the information protection by reducing the E2E-PLR as expected, but at the cost of the information rate; while higher values of $N$ increase protection at the price of complexity as well as encoding and decoding delay, which was not addressed for space reasons. Besides this numerical analysis, other optimization methods such as deep learning can also be used, as studied in \cite{machumilane2023learning}, where RL is used to determine the optimal PD rates at a mobile UE in DC with LEO satellites at different elevation angles.

\begin{table*}

\centering
\caption{Model Performance with Satellite $S1$ at $70^{\circ}$ and $S2$ at $60^{\circ}$. \\$LB = 1$ means that $N$ packets are scheduled on $S1$ only, $DT=0$ means PS only, and $DT=1$ means PD-only} 
\label{tab:evaluation_70_60}
\resizebox{0.9\textwidth}{!}{
\label{table_1}
\setlength{\tabcolsep}{3pt}
\begin{tabular}{|p{70pt} p{50pt} p{50pt} p{50pt} p{50pt} p{50pt} p{50pt}|p{70pt} p{50pt} p{50pt} p{50pt} p{50pt} p{50pt} p{50pt}|}
\hline

&\multicolumn{6}{c|}{\textbf{Packet Duplication + Packet Splitting}} &\multicolumn{7}{c|}{\textbf{Network Coding}}\\	
\hline
&\multicolumn{13}{c|}{}  \\
&\multicolumn{13}{c|}{\textbf{N=10}}  \\

\hline
 \textbf{LB on Sat A}&0.&	 0.2&	 0.4&	 0.6&	 0.8&	\textbf{1}& \textbf{LB on Sat A}&		 0.&	 0.2&	 0.4&	 0.6&	 0.8&	 1\\	
 \textbf{DT Ratio}&0.&	 0.&	 0.&	 0.&	 0.&	 \textbf{0}&		\textbf{Info Pkts (K)}& 2&	 2&	 2&	 2&	 2&	 2\\
 \textbf{RF}&1.&	 1.&	 1.&	 1.&	 1.&	\textbf{1}&		\textbf{Coding Rate} &5.&	 5.&	 5.&	 5.&	 5.&	 5\\	
 \textbf{E2E-PLR}&0.09311&	 0.0755282&	 0.0579464&	 0.0403645&	 0.0227827&	 \textbf{0.00520084} &\textbf{E2E-PLR}&		 0.0000954391&	 0.0000018026&	 0.000004103&	 0.000009&	 0.0000210529&	 0.000112862\\

 \hline
													
\textbf{LB on Sat A}& 0.&	 0.2&	 0.4&	 0.6&	 0.8&	 \textbf{1.}&	\textbf{LB on Sat A}&		 0.&	 0.2&	 0.4&	 0.6&	 0.8&	 1\\	
\textbf{DT Ratio}& 0.2&	 0.2&	 0.2&	 0.2&	 0.2&	\textbf{0.2}&		\textbf{Info Pkts (K)}& 4&	 4&	 4&	 4&	 4&	 4\\
 \textbf{RF}&1.2&	 1.2&	 1.2&	 1.2&	 1.2&	\textbf{1.2}&		\textbf{Coding Rate} & 2.5&	 2.5&	 2.5&	 2.5&	 2.5&	 2.5\\	
 \textbf{E2E-PLR}&0.0745849&	 0.0605194&	 0.0464539&	 0.0323885&	 0.018323&	\textbf{0.00425752}&	\textbf{E2E-PLR}&	 0.00179892&	 0.000825826&	 0.0000778478&	 0.000159476&	 0.0004734&	 0.000593578\\

 \hline
													
 \textbf{LB on Sat A}&0.&	 0.2&	 0.4&	 0.6&	 0.8&	 1.&	\textbf{LB on Sat A}&		 0.&	 0.2&	\textbf{0.4}&	 0.6&	 0.8&	 1\\	
 \textbf{DT Ratio}&0.4&	 0.4&	 0.4&	 0.4&	 0.4&	 0.4&		\textbf{Info Pkts (K)}& 6&	 6&	 \textbf{6}&	 6&	 6&	 6\\
 \textbf{RF}&1.4&	 1.4&	 1.4&	 1.4&	 1.4&	 1.4&	\textbf{Coding Rate} &	 1.66667&	 1.66667&	 \textbf{1.66667}&	 1.66667& 1.66667&	 1.66667\\
 \textbf{E2E-PLR}&0.0560597&	 0.0455106&	 0.0349615&	 0.0244124&	 0.0138633&	 0.00331421&	\textbf{E2E-PLR}&	 0.0165817&	 0.0104229&	 \textbf{0.0055983}&	 0.00152069&	 0.00183383&	 0.00213061\\

\hline
													
\textbf{LB on Sat A}& 0.&	 0.2&	 0.4&	 0.6&	 0.8&	 1.&	\textbf{LB on Sat A}&		 0.&	 0.2&	 0.4&	 0.6&	 0.8&	 1\\	
\textbf{DT Ratio}& 0.6&	 0.6&	 0.6&	 0.6&	 0.6&	 0.6&		\textbf{Info Pkts (K)}& 8&	 8&	 8&	 8&	 8&	 8\\
 \textbf{RF}&1.6&	 1.6&	 1.6&	 1.6&	 1.6&	 1.6&	\textbf{Coding Rate} &	 1.25&	 1.25&	 1.25&	 1.25&	 1.25&	 1.25\\	
\textbf{E2E-PLR}& 0.0375346&	 0.0305018&	 0.0234691&	 0.0164364&	 0.00940362&	 0.00237089&	\textbf{E2E-PLR}&	 0.101699&	 0.0766988&	 0.0541017&	 0.032141&	 0.00581474&	 0.00645361\\

\hline
													
\textbf{LB on Sat A}& 0.&	 0.2&	 0.4&	 0.6&\textbf{0.8}&	 1.&\textbf{LB on Sat A}&		 0.&	 0.2&	 0.4&	 0.6&	 0.8&	 1\\	
\textbf{DT Ratio}& 0.8&	 0.8&	 0.8&	 0.8&	\textbf{0.8}&	 0.8&\textbf{Info Pkts (K)}& 10&	 10&	 10&	 10&	 10&	 10\\
 \textbf{RF}&1.8&	 1.8&	 1.8&	 1.8&\textbf{1.8}&	 1.8&\textbf{Coding Rate} &	 1.&	 1.&	 1.&	 1.&	 1.&	 1\\	
 \textbf{E2E-PLR}&0.0190094&	 0.015493&	 0.0119767&	 0.0084603&\textbf{0.00494394}&	 0.00142757&\textbf{E2E-PLR}&	 0.449649&	 0.38909&	 0.319265&	 0.241459&	 0.15476&	 0.0175185\\

\hline
													
\textbf{LB on Sat A}&0.&	 0.2&	 0.4&	 0.6&	 0.8&	 1&\multicolumn{7}{c|}{} \\								
 \textbf{DT Ratio}&1.&	 1.&	 1.&	 1.&	 1.&	 1.&	\multicolumn{7}{c|}{} \\							
\textbf{RF}& 2.&	 2.&	 2.&	 2.&	 2.&	 2.&	\multicolumn{7}{c|}{} \\							
\textbf{E2E-PLR}& 0.000484251&	 0.000484251&	 0.000484251&	 0.000484251&	 0.000484251&	 0.000484251&	\multicolumn{7}{c|}{} \\

\hline
&\multicolumn{13}{c|}{}  \\
&\multicolumn{13}{c|}{\textbf{N=20}}  \\

\hline

  \textbf{LB on Sat A}& 0.&	 0.25&	 0.5&	 0.75&	 \textbf{1}& &	\textbf{LB on Sat A}&	 0.&	 0.25&	 0.5&	 0.75&	 1.&\\
\textbf{DT Ratio}&  0.&	 0.&	 0.&	 0.&	 \textbf{0}&			& \textbf{Info Pkts (K)}& 2&	 2&	 2&	 2&	 2&\\
\textbf{RF}& 1.&	 1.&	 1.&	 1.&	 \textbf{1}&			 &\textbf{Coding Rate} &10.&	 10.&	 10.&	 10.&	 10.&\\
 \textbf{E2E-PLR}& 0.09311&	 0.0711327&	 0.0491554&	 0.0271781&	 \textbf{0.00520084}&	&\textbf{E2E-PLR}& 		0.00000003562&	 0.00000000229&	 0.00000001828&	 0.0000001432&	 0.000002620&\\
 \hline
											
 \textbf{LB on Sat A}& 0.&	 0.25&	 0.5&	 0.75&	\textbf{1}&		&	\textbf{LB on Sat A}& 0.&	 0.25&	 0.5&	 0.75&	 1.&\\
 \textbf{DT Ratio}& 0.25&	 0.25&	 0.25&	 0.25&	 \textbf{0.25}&		&\textbf{Info Pkts (K)}&	 6&	 6&	 6&	 6&	 6&\\
\textbf{RF}& 1.25&	 1.25&	 1.25&	 1.25&	\textbf{1.25}& &\textbf{Coding Rate}  & 3.33333&	 3.33333&	 3.33333&	 3.33333&	 3.33333&\\
 \textbf{E2E-PLR}&  0.0745849&	 0.057003&	 0.0394212&	 0.0218394&	 \textbf{0.00425752}&	&	\textbf{E2E-PLR}& 	 0.0000134435&	0.00000161&	0.000004453&	 0.0000312472&	 0.0000500809&\\

 \hline
											
  \textbf{LB on Sat A}&0.&	 0.25&	 0.5&	 0.75&	 1.&		&\textbf{LB on Sat A}	& 0.&	 0.25&	 0.5&	 0.75&	 1.&\\
 \textbf{DT Ratio}& 0.5&	 0.5&	 0.5&	 0.5&	 0.5&			&\textbf{Info Pkts (K)}& 10&	 10&	 10&	 10&	 10&\\
\textbf{RF}& 1.5&	 1.5&	 1.5&	 1.5&	 1.5&			&\textbf{Coding Rate} & 2.&	 2.&	 2.&	 2.&	 2.&\\
\textbf{E2E-PLR}&  0.0560597&	 0.0428733&	 0.029687&	 0.0165006&	 0.00331421& &\textbf{E2E-PLR}& 	 0.000844959&	 0.000261506&	 0.000169146&	 0.000335702&	 0.000416254&\\

\hline
											
  \textbf{LB on Sat A}&0.&	 0.25&	 0.5&	 0.75&	 1.&		&\textbf{LB on Sat A}	& 0.&	 0.25&	\textbf{0.5}&	 0.75&	 1.&\\
 \textbf{DT Ratio}& 0.75&	 0.75&	 0.75&	 0.75&	 0.75&		&\textbf{Info Pkts (K)}&	 14&	 14&	\textbf{14}&	 14&	 14&\\
 \textbf{RF}&1.75&	 1.75&	 1.75&	 1.75&	 1.75&			&\textbf{Coding Rate} & 1.42857&	 1.42857&	 \textbf{1.42857}&	 1.42857&	 1.42857&\\
\textbf{E2E-PLR}&  0.0375346&	 0.0287436&	 0.0199527&	 0.0111618&	 0.00237089&	&\textbf{E2E-PLR}& 		 0.0219808&	 0.0111305&	 \textbf{0.005004}&	 0.00234712&	 0.00259776&\\

\hline
											
  \textbf{LB on Sat A}&0.&	 0.25&	 0.5&	 0.75&	 1.&		&\textbf{LB on Sat A}	& 0.&	 0.25&	 0.5&	 0.75&	 1.&\\
 \textbf{DT Ratio}& 1.&	 1.&	 1.&	 1.&	 1.&			&\textbf{Info Pkts (K)}& 18&	 18&	 18&	 18&	 18&\\
\textbf{RF}& 2.&	 2.&	 2.&	 2.&	 2.&			&\textbf{Coding Rate} & 1.11111&	 1.11111&	 1.11111&	 1.11111&	 1.11111&\\
\textbf{E2E-PLR}&  0.0190094&	 0.0146139&	 0.0102185&	 0.00582303&	 0.00142757& &\textbf{E2E-PLR}& 		 0.270414&	 0.19606&	 0.122455&	 0.0553938&	 0.0139531&\\
\hline
\end{tabular}
}
\label{evaluation}
\end{table*}
\begin{table*}
\centering
\caption{ Model Performance with Satellite $S1$ at $60^{\circ}$ and $S2$ at $45^{\circ}$}\label{tab:trans_probs_60_45}
\resizebox{0.9\textwidth}{!}{
\label{table_2}
\setlength{\tabcolsep}{3pt}
\begin{tabular}{|p{70pt} p{50pt} p{50pt} p{50pt} p{50pt} p{50pt} p{50pt}|p{70pt} p{50pt} p{50pt} p{50pt} p{50pt} p{50pt} p{50pt}|}
\hline

&\multicolumn{6}{c|}{\textbf{Packet Duplication + Packet Splitting}} &\multicolumn{7}{c|}{\textbf{Network Coding}}\\	
\hline
&\multicolumn{13}{c|}{}  \\
&\multicolumn{13}{c|}{\textbf{N=10}}  \\

\hline
\textbf{LB on Sat A}& 0.&	 0.2&	 0.4&	 0.6&	 0.8&	 1.&	\textbf{LB on Sat A}&	 0.&	 0.2&	 0.4&	 0.6&0.8&	 1\\
\textbf{DT Ratio}&  0.&	 0.&	 0.&	 0.&	 0.&	 0.&		\textbf{Info Pkts (K)}& 2&	 2&	 2&	 2&	 2&	 2\\
 \textbf{RF}& 1.&	 1.&	 1.&	 1.&	 1.&	 1.&		\textbf{Coding Rate} & 5.&	 5.&	 5.&	 5.&	 5.&	 5\\
\textbf{E2E-PLR}& 0.271081&	 0.235487&	 0.199893&	 0.164298&	 0.128704&	 0.09311&	\textbf{E2E-PLR}&	 0.0261343&	 0.00392668&	 0.00117048&	 0.000347407&	 0.000102725&	 0.0000954391\\

\hline
												
\textbf{LB on Sat A}& 0.&	 0.2&	 0.4&	 0.6&	 0.8&	 1.&	\textbf{LB on Sat A}&	 0.&	 0.2&	 0.4&	 \textbf{0.6}&	\textbf{0.8}&	 1\\
\textbf{DT Ratio}&  0.2&	 0.2&	 0.2&	 0.2&	 0.2&	 0.2&		\textbf{Info Pkts (K)}& 4&	 4&	 4&	\textbf{4}&	 \textbf{4}&	 4\\
\textbf{RF}&  1.2&	 1.2&	 1.2&	 1.2&	 1.2&	 1.2&		\textbf{Coding Rate} & 2.5&	 2.5&	 2.5&	 \textbf{2.5}&	 \textbf{2.5}&	 2.5\\
 \textbf{E2E-PLR}&0.221913&	 0.193437&	 0.164962&	 0.136487&	 0.108011&	 0.0795361&	\textbf{E2E-PLR}&	 0.0784484&	 0.0670728&	 0.0170166&	 \textbf{0.005662}&	 \textbf{0.00235425}&	 0.00179892\\

 \hline
												
 \textbf{LB on Sat A}&0.&	 0.2&	 0.4&	 0.6&	 0.8&	 1.&	\textbf{LB on Sat A}&	 0.&	 0.2&	 0.4&	 0.6&	 0.8&	 1\\
\textbf{DT Ratio}& 0.4&	 0.4&	 0.4&	 0.4&	 0.4&	 0.4&		\textbf{Info Pkts (K)}& 6&	 6&	 6&	 6&	 6&	 6\\
\textbf{RF}&  1.4&	 1.4&	 1.4&	 1.4&	 1.4&	 1.4&		\textbf{Coding Rate} & 1.66667&	 1.66667&	 1.66667&	 1.66667&	 1.66667&	 1.66667\\
\textbf{E2E-PLR}& 0.172745&	 0.151388&	 0.130032&	 0.108675&	 0.0873186&	 0.0659622&	\textbf{E2E-PLR}&	 0.166087&	 0.147993&	 0.128773&	 0.0470157&	 0.0223398&	 0.0165817\\

\hline
												
\textbf{LB on Sat A}& 0.&	 0.2&	 0.4&	 0.6&	 0.8&	 1.&		\textbf{LB on Sat A}& 0.&	 0.2&	 0.4&	 0.6&	 0.8&	 1\\
 \textbf{DT Ratio}& 0.6&	 0.6&	 0.6&	 0.6&	 0.6&	 0.6&		\textbf{Info Pkts (K)}& 8&	 8&	 8&	 8&	 8&	 8\\
 \textbf{RF}& 1.6&	 1.6&	 1.6&	 1.6&	 1.6&	 1.6&		 \textbf{Coding Rate} &1.25&	 1.25&	 1.25&	 1.25&	 1.25&	 1.25\\
\textbf{PLR}& 0.123577&	 0.109339&	 0.0951012&	 0.0808635&	 0.0666259&	 0.0523882&		\textbf{PLR}& 0.298515&	 0.273982&	 0.261917&	 0.246764&	 0.136753&	 0.101699\\

\hline
												
 \textbf{LB on Sat A}&0.&	 0.2&	 0.4&	 0.6&	 0.8&	 1.&	\textbf{LB on Sat A}&	 0.&	 0.2&	 0.4&	 0.6&	 0.8&	 1\\
\textbf{DT Ratio}&  0.8&	 0.8&	 0.8&	 0.8&	 0.8&	 0.8&		\textbf{Info Pkts (K)}& 10&	 10&	 10&	 10&	 10&	 10\\
\textbf{RF}&  1.8&	 1.8&	 1.8&	 1.8&	 1.8&	 1.8&		\textbf{Coding Rate} & 1.&	 1.&	 1.&	 1.&	 1.&	 1\\
\textbf{E2E-PLR}& 0.0744084&	 0.0672896&	 0.0601708&	 0.0530519&	 0.0459331&	 0.0388143&		\textbf{E2E-PLR}& 0.4846&	 0.522416&	 0.538365&	 0.553782&	 0.568683&	 0.449649\\

\hline
												
 \textbf{LB on Sat A}&0.&	 0.2&	 0.4&	 0.6&	 0.8&	 1.&	\multicolumn{7}{c|}{} \\							
\textbf{DT Ratio}&  1.&	 1.&	 1.&	 1.&	 1.&	 1.&\multicolumn{7}{c|}{} \\								
\textbf{RF}&  2.&	 2.&	 2.&	 2.&	 2.&	 2.&	\multicolumn{7}{c|}{} \\							
\textbf{E2E-PLR}& 0.0252403&	 0.0252403&	 0.0252403&	 0.0252403&	 0.0252403&	 0.0252403&	\multicolumn{7}{c|}{} \\	

\hline

	&\multicolumn{13}{c|}{}  \\
&\multicolumn{13}{c|}{\textbf{N=20}}  \\

\hline
\textbf{LB on Sat A}& 0.&	 0.25&	 0.5&	 0.75&	 1.&	&\textbf{LB on Sat A}&	 0.&	 0.25&	 0.5&	 0.75&	 1.&\\		
 \textbf{DT Ratio}& 0.&	 0.&	 0.&	 0.&	 0.&	&	 \textbf{Info Pkts (K)}&2&	 2&	 2&	 2&	 2&		\\
 \textbf{RF}& 1.&	 1.&	 1.&	 1.&	 1.&		&\textbf{Coding Rate} & 10.&	 10.&	 10.&	 10.&	 10.&\\		
 \textbf{E2E-PLR}&0.271081&	 0.226588&	 0.182095&	 0.137603&	 0.09311&		&\textbf{E2E-PLR}& 0.0046447&	 0.000110631&	 0.000005240&	 0.0000002437&	 0.0000000356&		\\

 \hline
												
 \textbf{LB on Sat A}&0.&	 0.25&	 0.5&	 0.75&	 1.&		&\textbf{LB on Sat A}& 0.&	 0.25&	 0.5&	 0.75&	 1.&	\\	
\textbf{DT Ratio}&  0.2&	 0.2&	 0.2&	 0.2&	 0.2& &\textbf{Info Pkts (K)}& 6&	 6&	 6&	 6&	 6&		\\
\textbf{RF}&   1.2&	 1.2&	 1.2&	 1.2&	 1.2& &\textbf{Coding Rate} &	 3.33333 & 3.33333&	 3.33333&	 3.33333&	 3.33333&\\		
\textbf{E2E-PLR}&   0.221913&	 0.186319&	 0.150724&	 0.11513&	 0.0795361& &\textbf{E2E-PLR}& 0.0330652&	 0.0208113&	 0.00101701&	 0.0000709973&	 0.0000134435&		\\

\hline
												
\textbf{LB on Sat A}& 0.&	 0.25&	 0.5&	 0.75&	 1.&		&\textbf{LB on Sat A}& 0.&	 0.25&	 0.5&	 \textbf{0.75}&	 1.&	\\	
\textbf{DT Ratio}&  0.4&	 0.4&	 0.4&	 0.4&	 0.4& &\textbf{Info Pkts (K)}& 10&	 10&	 10&	 \textbf{10}&	 10&	\\	
\textbf{RF}&   1.4&	 1.4&	 1.4&	 1.4&	 1.4&	 &\textbf{Coding Rate}&2.&	 2.&	 2.&	 \textbf{2}&	 2.&		\\
\textbf{PLR}& 0.172745&	 0.146049&	 0.119353&	 0.0926578&	 0.0659622&	&\textbf{E2E-PLR}&	 0.105922&	 0.0802589&	 0.0305701&	\textbf{0.00346057}&	 0.000844959&		\\

 \hline
												
\textbf{LB on Sat A}& 0.&	 0.25&	 0.5&	 0.75&	 1.&		&\textbf{LB on Sat A}& 0.&	 0.25&	 0.5&	 0.75&	 1.&	\\	
\textbf{DT Ratio}&  0.6&	 0.6&	 0.6&	 0.6&	 0.6&	&\textbf{Info Pkts (K)}& 14&	 14&	 14&	 14&	 14&	\\	
\textbf{RF}&  1.6&	 1.6&	 1.6&	 1.6&	 1.6&		&\textbf{Coding Rate} & 1.42857&	 1.42857&	 1.42857&	 1.42857&	 1.42857&	\\	
\textbf{E2E-PLR}&  0.123577&	 0.105779&	 0.0879824&	 0.0701853&	 0.0523882&		&\textbf{E2E-PLR}& 0.249758&	 0.210091&	 0.166082&	 0.0651273&	 0.0219808&\\
\hline
												
\textbf{LB on Sat A}& 0.&	 0.25&	 0.5&	 0.75&	 1.&		&\textbf{LB on Sat A}& 0.&	 0.25&	 0.5&	 0.75&	 1.&	\\	
 \textbf{DT Ratio}&  0.8&	 0.8&	 0.8&	 0.8&	 0.8&		&\textbf{Info Pkts (K)}& 18&	 18&	 18&	 18&	 18&		\\
\textbf{RF}&   1.8&	 1.8&	 1.8&	 1.8&	 1.8&		&\textbf{Coding Rate} & 1.11111&	 1.11111&	 1.11111&	 1.11111&	 1.11111&		\\
\textbf{E2E-PLR}& 0.0744084&	 0.0655099&	 0.0566114&	 0.0477128&	 0.0388143&	&\textbf{E2E-PLR}& 0.489212&	 0.456393&	 0.429404&	 0.405293&	 0.270414&\\

\hline
\end{tabular}
}
    
\end{table*}

\begin{table*}
\centering
\caption{ Model Performance with Satellite $S1$ at $70^{\circ}$ and $S2$ at $45^{\circ}$.} \label{tab:trans_probs_70_45}
\resizebox{0.9\textwidth}{!}{
\label{table_3}
\setlength{\tabcolsep}{3pt}
\begin{tabular}{|p{70pt} p{50pt} p{50pt} p{50pt} p{50pt} p{50pt} p{50pt}|p{70pt} p{50pt} p{50pt} p{50pt} p{50pt} p{50pt} p{50pt}|}
\hline
&\multicolumn{6}{c|}{\textbf{Packet Duplication + Packet Splitting}} &\multicolumn{7}{c|}{\textbf{Network Coding}}\\	
\hline
&\multicolumn{13}{c|}{}  \\
&\multicolumn{13}{c|}{\textbf{N=10}}  \\

\hline
\textbf{LB on Sat A}& 0.&	 0.2&	 0.4&	 0.6&	 0.8&	 \textbf{1}&	\textbf{LB on Sat A}&	 0.&	 0.2&	 0.4&	 0.6&	 0.8&	 1\\
 \textbf{DT Ratio}&  0.&	 0&	 0.&	 0.&	 0.&	 \textbf{0}&\textbf{Info Pkts (K)}&		 2&	 2&	 2&	 2&	 2&	 2\\
\textbf{RF}&  1.&	1 &	 1.&	 1.&	 1.&	\textbf{1}&	\textbf{Coding Rate} 	& 5.&	 5.&	 5.&	 5.&	 5.&	 5\\
\textbf{E2E-PLR}& 0.271081&	0.217905&	 0.164729&	 0.111553&	 0.0583768&	 \textbf{0.00520084}&	\textbf{E2E-PLR}&	 0.0261343&	 0.000193402&	 0.000129046&	 0.0000861034&	 0.0000574499&	 0.000112862\\

\hline
												
 \textbf{LB on Sat A}&0.&	 0.2&	 0.4&	 0.6&	 0.8&	 \textbf{1}.&		\textbf{LB on Sat A}& 0.&	 0.2&	 0.4&	 0.6&	 0.8&	 1\\
 \textbf{DT Ratio}&  0.2&	 0.2&	 0.2&	 0.2&	 0.2&	 \textbf{0.2}&		\textbf{Info Pkts (K)}& 4&	 4&	 4&	 4&	 4&	 4\\
\textbf{RF}& 1.2&	 1.2&	 1.2&	 1.2&	 1.2&	 \textbf{1.2}&		 \textbf{Coding Rate} &2.5&	 2.5&	 2.5&	 2.5&	 2.5&	 2.5\\
\textbf{E2E-PLR}& 0.217147&	 0.174606&	 0.132065&	 0.0895242&	 0.0469834&	 \textbf{0.00444264}&	\textbf{E2E-PLR}&	 0.0784484&	 0.0644295&	 0.000784292&	 0.000520812&	 0.00059709&	 0.000593578\\

\hline
												
\textbf{LB on Sat A}& 0.&	 0.2&	 0.4&	 0.6&	 0.8&	 1.&	\textbf{LB on Sat A}&	 0.&	 0.2&	 0.4&	\textbf{0.6}&	 0.8&	 1\\
\textbf{DT Ratio}&  0.4&	 0.4&	 0.4&	 0.4&	 0.4&	 0.4&		\textbf{Info Pkts (K)}& 6&	 6&	 6&	 \textbf{6}&	 6&	 6\\
\textbf{RF}&  1.4&	 1.4&	 1.4&	 1.4&	 1.4&	 1.4&	\textbf{Coding Rate} &	 1.66667&	 1.66667&	 1.66667&	 \textbf{1.66667}&	 1.66667&	 1.66667\\
\textbf{E2E-PLR}& 0.163212&	 0.131307&	 0.0994012&	 0.0674956&	 0.03559&	 0.00368445&	\textbf{E2E-PLR}&	 0.166087&	 0.143144&	 0.119131&	 \textbf{0.00265513}&	 0.00224985&	 0.00213061\\

\hline
												
\textbf{LB on Sat A}& 0.&	 0.2&	 0.4&	 0.6&	 0.8&	 1.&		\textbf{LB on Sat A}& 0.&	 0.2&	 0.4&	 0.6&	 0.8&	 1\\
 \textbf{DT Ratio}&  0.6&	 0.6&	 0.6&	 0.6&	 0.6&	 0.6&		\textbf{Info Pkts (K)}& 8&	 8&	 8&	 8&	 8&	 8\\
\textbf{RF}&  1.6&	 1.6&	 1.6&	 1.6&	 1.6&	 1.6&		\textbf{Coding Rate} & 1.25&	 1.25&	 1.25&	 1.25&	 1.25&	 1.25\\
\textbf{E2E-PLR}& 0.109278&	 0.0880078&	 0.0667374&	 0.045467&	 0.0241966&	 0.00292625&	\textbf{E2E-PLR}&	 0.298515&	 0.266332&	 0.233561&	 0.197813&	 0.0070158&	 0.00645361\\

\hline										
\textbf{LB on Sat A}& 0.&	 0.2&	 0.4&	 0.6&	 0.8&	 1.&		\textbf{LB on Sat A}& 0.&	 0.2&	 0.4&	 0.6&	 0.8&	 1\\
  \textbf{DT Ratio}& 0.8&	 0.8&	 0.8&	 0.8&	 0.8&	 0.8&		 \textbf{Info Pkts (K)}&10&	 10&	 10&	 10&	 10&	 10\\
\textbf{RF}&  1.8&	 1.8&	 1.8&	 1.8&	 1.8&	 1.8&		\textbf{Coding Rate} & 1.&	 1.&	 1.&	 1.&	 1.&	 1.\\
\textbf{E2E-PLR}& 0.055344&	 0.0447088&	 0.0340736&	 0.0234384&	 0.0128032&	 0.00216805&	\textbf{E2E-PLR}&	 0.4846&	 0.446993&	 0.404365&	 0.35845&	 0.308997&	 0.0175185\\

\hline
							
\textbf{LB on Sat A}& 0.&	 0.2&	 0.4&	 0.6&	 0.8&	 1.&	\multicolumn{7}{c|}{} \\						
 \textbf{DT Ratio}&  1.&	 1.&	 1.&	 1.&	 1.&	 1.&	\multicolumn{7}{c|}{} \\						
 \textbf{RF}& 2.&	 2.&	 2.&	 2.&	 2.&	 2.&			\multicolumn{7}{c|}{} \\				
\textbf{E2E-PLR}& 0.00140985&	 0.00140985&	 0.00140985&	 0.00140985&	 0.00140985&	 0.00140985&		\multicolumn{7}{c|}{} \\

\hline
&\multicolumn{13}{c|}{}  \\
&\multicolumn{13}{c|}{\textbf{N=20}}  \\

    \hline
												
\textbf{LB on Sat A}& 0.&	 0.25&	 0.5&	 0.75&	\textbf{1}&	&		\textbf{LB on Sat A}& 0.&	 0.25&	 0.5&	 0.75&	 1.&	\\
 \textbf{DT Ratio}&  0.&	 0.&	 0.&	 0.&	 \textbf{0}&			 &\textbf{Info Pkts (K)}& 2&	 2&	 2&	 2&	 2&	\\
\textbf{RF}&  1.&	 1.&	 1.&	 1.&	 \textbf{1}&			&\textbf{Coding Rate}& 10.&	 10.&	 10.&	 10.&	 10.&\\	
\textbf{E2E-PLR}& 0.271081&	 0.204611&	 0.138141&	 0.0716708&	 \textbf{0.00520084}&		&\textbf{E2E-PLR}&	 0.0046447&	 0.0000184924&	 0.000006726&	 0.000002446&	 0.000002620&	\\

\hline
												
\textbf{LB on Sat A}& 0.&	 0.25&	 0.5&	 0.75&	 \textbf{1}&		&	\textbf{LB on Sat A}& 0.&	 0.25&	 0.5&	 0.75&	 1.&	\\
  \textbf{DT Ratio}& 0.2&	 0.2&	 0.2&	 0.2&	 \textbf{0.2}&			&\textbf{Info Pkts (K)}& 6&	 6&	 6&	 6&	 6&	\\
\textbf{RF}&  1.2&	 1.2&	 1.2&	 1.2&	 \textbf{1.2}&			&\textbf{Coding Rate} & 3.33333&	 3.33333&	 3.33333&	 3.33333&	 3.33333&\\	
\textbf{E2E-PLR}& 0.217147&	 0.163971&	 0.110795&	 0.0576186&	 \textbf{0.00444264}&	&	\textbf{E2E-PLR}&	 0.0330652&	 0.0183088&	 0.00016312&	 0.0000701162&	 0.0000500809&	\\

\hline
												
\textbf{LB on Sat A}& 0.&	 0.25&	 0.5&	 0.75&	 1.&		&	\textbf{LB on Sat A}& 0.&	 0.25&	 0.5&	 0.75&	 1.&	\\
 \textbf{DT Ratio}&  0.4&	 0.4&	 0.4&	 0.4&	 0.4&			&\textbf{Info Pkts (K)}& 10&	 10&	 10&	 10&	 10&	\\
 \textbf{RF}& 1.4&	 1.4&	 1.4&	 1.4&	 1.4&			&\textbf{Coding Rate}& 2.&	 2.&	 2.&	 2.&	 2.&	\\
\textbf{E2E-PLR}& 0.163212&	 0.12333&	 0.0834484&	 0.0435664&	 0.00368445&	&	\textbf{E2E-PLR}&	 0.105922&	 0.0735551&	 0.00148427&	 0.000642388&	 0.000416254&	\\

\hline
												
\textbf{LB on Sat A}& 0.&	 0.25&	 0.5&	 0.75&	 1.&		&\textbf{LB on Sat A}&	 0.&	 0.25&	 0.5&	 \textbf{0.75}&	1&	\\
 \textbf{DT Ratio}&  0.6&	 0.6&	 0.6&	 0.6&	 0.6&			&\textbf{Info Pkts (K)}& 14&	 14&	 14&	 \textbf{14}&	 14&	\\
\textbf{RF}&  1.6&	 1.6&	 1.6&	 1.6&	 1.6&			&\textbf{Coding Rate}& 1.42857&	 1.42857&	 1.42857&	 \textbf{1.42857}&	 1.42857&\\	
\textbf{E2E-PLR}& 0.109278&	 0.0826902&	 0.0561022&	 0.0295142&	 0.00292625&		&\textbf{E2E-PLR}&	 0.249758&	 0.196745&	 0.139319&	 \textbf{0.00420139}&	 0.00259776&	\\

\hline
												
\textbf{LB on Sat A}& 0.&	 0.25&	 0.5&	 0.75&	 1.&	&	\textbf{LB on Sat A}&	 0.&	 0.25&	 0.5&	 0.75&	 1.&	\\
 \textbf{DT Ratio}&  0.8&	 0.8&	 0.8&	 0.8&	 0.8&			&\textbf{Info Pkts (K)}& 18&	 18&	 18&	 18&	 18&	\\
 \textbf{RF}& 1.8&	 1.8&	 1.8&	 1.8&	 1.8&			&\textbf{Coding Rate}& 1.11111&	 1.11111&	 1.11111&	 1.11111&	 1.11111&\\	
\textbf{E2E-PLR}& 0.055344&	 0.04205&	 0.028756&	 0.015462&	 0.00216805&	&	\textbf{E2E-PLR}&	 0.489212&	 0.420909&	 0.341024&	 0.248288&	 0.0139531&	\\

\hline

\end{tabular}
}
\end{table*}

\section{Conclusion}
This paper introduces the mathematical framework for calculating the average end-to-end packet loss in case of loss processes modeled through a DMC when combining PD, PS or when NC is employed in 5G-DC. Such models are valuable metrics for designing optimal policies with full knowledge of the loss process, which can be compared to empirical models learned through machine learning algorithms.
Evaluating the model for DC to LEO satellites reveals interesting insights for designing efficient and reliable communications. We analyzed the effectiveness of PD, PS and NC in mitigating the end-to-end packet loss rate. Although simulation results show the effectiveness of NC in reducing end-to-end packet loss, one must consider the increased complexity introduced by the encoding/decoding procedure, which affects information rate by introducing encoding/decoding delays, as extensively studied in the RLNC literature. In challenging scenarios like $60^{\circ}$-$45^{\circ}$, NC outperforms PD and PS in achieving a target PLR, but introduces delay and computational overhead.

\section*{Acknowledgment}
This work was supported by the European Union under the Italian National Recovery and Resilience Plan (PNRR) PE00000001 - program "RESTART", and 
by the HORIZON-CL4-2021-SPACE-01 project "5G+ evoluTion to mutioRbitAl multibaNd neTwORks" (TRANTOR) No. 101081983


%





\ifCLASSOPTIONcaptionsoff
  \newpage
\fi




\vspace{2cm}
\bibliographystyle{IEEEtran}
\bibliography{Bibliography}

\vfill


\end{document}